\documentclass[10pt,final]{IEEEtran}

\usepackage{graphicx}
\usepackage[T1]{fontenc}
\usepackage{lmodern}
\usepackage{textcomp}
\usepackage{latexsym}
\usepackage{threeparttable} % 表での脚注の使用用
\usepackage{url} % 文献リスト用
\makeatletter
\def\Hline{%
\noalign{\ifnum0=`}\fi\hrule \@height 1.5pt \futurelet
\reserved@a\@xhline}
\makeatother

\begin{document}

\pagestyle{empty}
\clearpage

\title{Breakdown of a Benchmark Score Without\\
Internal Analysis of Benchmarking Program}
%\title{Towards a Breakdown of a Benchmark Score\\
%Without Internal Analysis of the Benchmarking Program}

\author{
\IEEEauthorblockN{Naoki Matagawa\IEEEauthorrefmark{1}, Kazuyuki Shudo}\\
\IEEEauthorblockA{Tokyo Institute of Technology}
%\thanks{This work was supported by JSPS KAKENHI Grant Numbers 25700008 and 26540161.}
}

\maketitle\thispagestyle{empty}

\begin{abstract}
A breakdown of a benchmark score is how much each aspect of the system performance affects the score.
Existing methods require internal analysis on the benchmarking program and then involve the following problems:
(1) require a certain amount of labor for code analysis, profiling, simulation, and so on and (2) require the benchmarking program itself.
In this paper, we present a method for breaking down a benchmark score without internal analysis of the benchmarking program.
The method utilizes regression analysis of benchmark scores on a number of systems.
Experimental results with 3 benchmarks on 15 Android smartphones showed that our method could break down those benchmark scores even though there is room for improvement in accuracy.
\end{abstract}


\footnote[0]{\IEEEauthorrefmark{1} Present affiliation is Asial Corporation}

\section{Introduction}

In this paper,
we present a method for breaking down a benchmark score without internal analysis of the benchmarking program.
The method quantifies a portion of a number of aspects of the system performance with regression analysis using number of systems.

%In this section, we present the summary of our research.

%\subsection{Concepts around benchmarking}

\textit{Benchmarking} is a qualitative or quantitative measurement of how well the program has been executed in a certain \textit{metric}.
A value obtained from benchmarking is called a \textit{benchmark score} or solely \textit{score}.
A program executed in benchmarking is called a \textit{benchmarking program} or solely \textit{program}.
The terms benchmarking, benchmark score, and benchmarking program are sometimes abbreviated to \textit{benchmark}.

%\subsection{Breakdown of benchmark score and motivation}

\begin{figure}[t]
    \centering
    \includegraphics[width=0.40\textwidth]{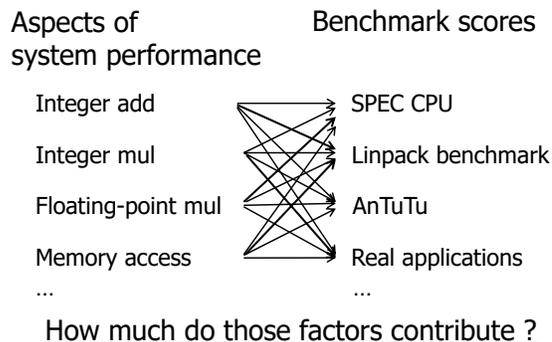}
    \caption{Research goal.}
    \label{fig:abstract}
\end{figure}

Our goal here is \textit{breakdown} of a benchmark score as shown in Figure \ref{fig:abstract}.
Breakdown of a benchmark score is an analysis on how much each aspect of the system performance affects the score.
Examples of those aspects are integer arithmetic and memory access.
Similar concepts such as performance modeling and execution time estimation are considered to be a kind of breakdown.
Breakdown of a benchmark score can be utilized for detecting bottlenecks, optimizing programs, and improving architectures \cite{GPUAnalysis}.
It can be also applied for identifying characteristics of a program, comparing systems, comparing programs quantitatively, and evaluating a bias of a system ranking.

%\subsection{Existing methods and problems}

Existing breakdown methods require internal analysis on the benchmarking program and then involve the following problems:
(1) require a certain amount of labor for code analysis, profiling, simulation, and so on and (2) require the benchmarking program itself.
%
%\subsection{Proposed method}
%
Our contribution presented in this paper is a breakdown method without those problems.
The proposed method breaks down a benchmark score without internal analysis of the benchmarking program.
It utilizes regression analysis of benchmark scores on a number of systems.

%\subsection{Experimental results}

To evaluate validity of the method,
we applied the method to 3 benchmarks on 15 Android smartphones.
The method could break down obtained benchmark scores even though there is room for improvement in accuracy.

%\subsection{Structure of this paper}

The rest of this paper is structured as follows.
Section {\bfseries \ref{sec:existing}} describes existing methods and their problems.
Section {\bfseries \ref{sec:proposed}} presents our method.% that solves the problems of the existing methods.
In Section {\bfseries \ref{sec:experiment}}, we evaluate validity of the method by applying the method to benchmarks.
Finally, we summarize our contribution and describe future work in Section {\bfseries \ref{sec:conclusion}}.

\section{Existing methods and problems}
\label{sec:existing}

Existing breakdown methods are based on
subprogram execution \cite{GPUAnalysis},
static code analysis \cite{SourceLevel},
execution trace analysis \cite{Seltzer},
profiling,
simulation and so on.

Without internal analysis, such as profiling, on a benchmarking program, any existing method cannot break down a benchmark score.
This is because existing methods run a target benchmark on the only one system
and then there is only one sample of relationships between aspects of system performance and the score.
Given only one sample, we can get no information about the relationships.

Existing methods require internal analysis on the benchmarking program and then involve the following problems:
(1) require a certain amount of labor for code analysis, profiling, simulation, and so on and (2) require the benchmarking program itself.
Especially, (2) is considered to be a critical problem because (2) makes it impossible to break down a benchmark score generated by a benchmark which restricts reverse engineering.

\section{Breakdown of a benchmark score by regression analysis}
\label{sec:proposed}

In this section, we describe a method for breaking down a benchmark score of arbitrary benchmark without internal analysis on the benchmarking program.

%\subsection{Use of regression analysis}

As we describe in Section {\bfseries \ref{sec:existing}},
we have to analyze inside of a benchmarking program
if there is only one sample of relationships between aspects of system performance and a benchmark score.
However, number of such samples enable regression analysis based on them
and we no longer need to analyze inside of the benchmarking program.
Our proposed method utilizes regression analysis,
and multiple linear regression analysis especially in the experiment presented in Section \ref{sec:experiment}.
%However, if there are many samples of the relationships, we no longer need to analyze inside of the benchmarking program
%because we can carry out regression analysis based on those samples.
%
%Therefore, our proposed method utilizes regression analysis.

%\subsection{Use of multiple systems} % 多数のシステムでのベンチマークスコアの測定の必要性

Explanatory variables for regression are merits in aspects of system performance
and a dependent variable is a benchmark score.
Examples of the aspects are integer arithmetic and memory access.
It should possible to estimate the merits in the aspects by running an adequate set of small benchmarks or \textit{microbenchmarks} \cite{Seltzer}.
We prepare such a set of microbenchmarks by ourselves for the experiment in Section \ref{sec:experiment}.
There should be a better set of microbenchmarks and its development or collection is an open problem.

It is possible to obtain multiple samples by running the same set of benchmarks on multiple systems.
Our proposed method premises that it is possible to obtain benchmark scores on multiple systems.
Published scores are available for widely-used benchmarks such as SPEC benchmarks,
and we can run benchmarking programs by ourselves of course.

The breakdown process described above is considered as confirmatory factor analysis (CFA).
The factors are merits in aspects of system performance.

\section{Evaluation}
\label{sec:experiment}

%Our proposed method outputs some breakdown results anyway.
In this section, we confirm that the proposed method is practical by applying the method to real benchmarks and systems.

%\subsection{Experimental Objective} % 実験目的は「実験目的」という見出しの下で述べた方が読者に実験目的をより確実に伝えることができると思うので，節を用意
%
%Although it is clear that our proposed method can break down a benchmark score without internal analysis on the benchmarking program,
%it remains to be seen whether the result of the breakdown is fully practical or not.
%Then, we performed an experiment to make it clear.

\subsection{Experiment}

In this experiment, we quantify aspects of system performance by developing and running microbenchmarks.
And then we break down scores of 3 benchmarks into the aspects by regression analysis.

First, we describe the following points which we must define to apply our method:
(i) target systems,
(ii) target benchmarks of breakdown,
(iii) microbenchmarks used to quantify each aspect of system performance.
After that, we describe other information about details of this experiment.

\subsubsection{Target systems}

We chose smartphones as target systems because smartphones meet the following two conditions:
(1) we can obtain enough samples by preparing number of systems which specification differs from each other by utilizing a certain remote service,
(2) such remote services do not utilize virtual machines to offer their service.
As a proof, we describe the characteristics of supercomputers, commodity computers, and smartphones.

We can hardly prepare number of supercomputers due to their availability.
Therefore supercomputers meet neither (1) nor (2).

In the case of commodity computers, tens of systems can be prepared through the remote services such as Amazon Elastic Compute Cloud (EC2) \cite{AmazonEC2} and Google Compute Engine \cite{GCE}.
Remote services of commodity computers, however, often utilize hypervisors such as Xen \cite{Xen} to offer their service; therefore the services are not suited for benchmarking.
This is because programs running on hypervisors contain unignorable overhead \cite{Overhead}.
Therefore commodity computers meet (1) but not (2).

In the case of smartphones, tens of systems can also be prepared through remote services such as Amazon Device Farm \cite{AmazonDeviceFarm}.
As of February 2016, Amazon Device Farm offers 18 iOS devices and 66 Android devices if we regard devices which differ from each other in its career or its OS version as a same device \footnote{https://aws.amazon.com/device-farm/device-list/}.
In addition, in contrast to remote services offering commodity computers, Amazon Device Farm do not utilize hypervisors to offer its service and then is suited to benchmarking since programs running on its service do not contain overhead.
Therefore smartphones meet both (1) and (2).

\vspace{1.5mm}

\noindent \textbf{Smartphones used in the experiment.}
In this experiment, we rented 15 Android devices shown in Table \ref{tab:androids} from Amazon Device Farm and experimented with them.

\begin{table*}[t]
	\centering
	\begin{threeparttable}[htb]
		\centering
		\caption{Android devices used in the experiment.}
		\label{tab:androids}
		\begin{tabular}{lp{5.5cm}p{1.5cm}l}
			\Hline
			{\small \textbf{Device name} (and career name)}    & {\small \textbf{Chipset}　(or processor)}\tnote{*1} & {\small \textbf{ISA}}\tnote{*1} & {\small \textbf{OS version}} \\
			\hline
			Amazon Fire Phone               & (unknown)                                 & A32          & 4.4.4 \\
			ASUS Memo Pad 7                 & (Intel Atom Z3745)                        & x86-64       & 5.0 \\
			ASUS Nexus 7 - 2nd Gen          & Qualcomm APQ8064                          & A32          & 6.0 \\
			HTC One M7 (AT\&T)              & (unknown)                                 & A32          & 4.4.2 \\
			HTC One M8 (AT\&T)              & Qualcomm MSM8974AB                        & A32          & 4.4.4 \\
			%HTC One M9 (AT\&T)              & Qualcomm MSM8994                          & A64          & 5.0.2 \\
			Huawei Nexus 6P                 & Qualcomm MSM8994                          & A64          & 6.0 \\
			LG G3 (AT\&T)                   & Qualcomm MSM8974AC                        & A32          & 5.0.1 \\
			LG G4 (Verizon)                 & Qualcomm MSM8992                          & A64          & 5.1 \\
			%LG Optimus L90 (T-Mobile)       & Qualcomm MSM8226                          & A32          & 4.4.2 \\
			Motorola Moto G (AT\&T)         & Qualcomm MSM8226                          & A32          & 4.4.4 \\
			%Motorola Moto X (Verizon)       & Qualcomm MSM8974AC                        & A32          & 5.1 \\
			Motorola Nexus 6                & Qualcomm APQ8084                          & A32          & 6.0 \\
			Samsung Galaxy Note 4 (AT\&T)   & Qualcomm APQ8084                          & A32          & 5.0.1 \\
			Samsung Galaxy Note 5           & Samsung Exynos 7420                       & A64          & 5.1.1 \\
			%Samsung Galaxy S3 (AT\&T)       & (unknown)                                 & A32          & 4.3 \\
			Samsung Galaxy S4 (AT\&T)       & (unknown)                                 & A32          & 5.0.1 \\
			Samsung Galaxy S6 Edge+ (AT\&T) & Samsung Exynos 7420                       & A64          & 5.1.1 \\
			Sony Xperia Z3 (GSM)            & (unknown)                                 & (unknown)    & 4.4.4 \\
			\Hline
		\end{tabular}
		\begin{tablenotes}
			\footnotesize
			\item[*1] Retrieved by executing \texttt{cat /proc/cpuinfo} on each device.
		\end{tablenotes}
	\end{threeparttable}
\end{table*}

\subsubsection{Target benchmarks}

SPEC CPU2006 \cite{CPU2006} is a widely-used benchmark suite to measure processor performance.
We prepared target benchmarks by porting a part of the SPEC CPU2006 benchmarks to Android devices.
As of February 2016, the newest version of SPEC CPU2006 is V1.2.
In this experiment, we use V1.1 because we could not obtain V1.2 due to no response from SPEC.
V1.2 improves compatibility, stability, documentation and ease of use and the results with V1.2 and V1.1 are still compatible.

SPEC CPU2006 is composed of 29 benchmarks.
In this experiment, however, we used the only benchmarks each of which is written in C due to portability issues.
Table \ref{tab:macrobenchmarks} shows the benchmarks we used and Table \ref{tab:environments} shows the environments and the compiler options with which we compiled the benchmarks.

Each of benchmarking programs in SPEC CPU2006 has three workloads; \texttt{train} (data for feedback-directed optimization), \texttt{test} (data for checking functionality), and \texttt{ref} (the real data set required for result reporting).
Because \texttt{ref} caused a memory allocation error in some Android devices, we used \texttt{test} and a custom workload based on \texttt{ref} in this experiment.

\begin{table}[t]
	\centering
	\begin{threeparttable}[htb]
		\centering
		\caption{Target benchmarks.}
		\label{tab:macrobenchmarks}
		\begin{tabular}{p{2.5cm}p{3.5cm}p{1.6cm}}
			\Hline
			\textbf{Name and abstract} & \textbf{Program details} & \textbf{Workload} \\
			\hline
			%400.perlbench \\

			%401.bzip2 \\

			%403.gcc \\

			\parbox[t]{2.5cm}{\textbf{429.mcf} Combinatorial Optimization} & \parbox[t]{3.5cm}{Vehicle scheduling. Uses a network simplex algorithm to schedule public transport.} & \texttt{test} \\ % Combinatorial Optimization
			% Vehicle scheduling. Uses a network simplex algorithm (which is also used in commercial products) to schedule public transport.

			%433.milc \\

			%445.gobmk \\

			%456.hmmer \\

			%458.sjeng \\
			\hline
			\parbox[t]{2.5cm}{\textbf{462.libquantum} Quantum Computing} & \parbox[t]{3.5cm}{Simulates a quantum computer, running Shor's polynomial-time factorization algorithm.} & \parbox[t]{1.6cm}{custom \tnote{*1}} \\ % Physics / Quantum Computing
			% Simulates a quantum computer, running Shor's polynomial-time factorization algorithm.

			%464.h264ref \\
			\hline
			\parbox[t]{2.5cm}{\textbf{470.lbm} Fluid Dynamics} & \parbox[t]{3.5cm}{Implements the "Lattice-Boltzmann Method" to simulate incompressible fluids in 3D.} & \texttt{test} \\ % Fluid Dynamics
			% Implements the "Lattice-Boltzmann Method" to simulate incompressible fluids in 3D

			%482.sphinx3 \\
			\Hline
		\end{tabular}
		\begin{tablenotes}
			\footnotesize
			\item[*1] 462.libquantum requires (a) size of problem and (b) random seed as input parameters.
			The workload \texttt{test} takes (a) 33 and (b) 5.
			The workload \texttt{ref} takes (a) 1397 and (b) 8.
			We used a custom workload which takes (a) 130 and (b) 8 because \texttt{test} workload resulted in too short execution time and \texttt{ref} workload resulted in too long execution time.
		\end{tablenotes}
	\end{threeparttable}
\end{table}

% Q. code は uncountable ？
In our implementation, there is a bit of native code embedded into the Android application which calls \texttt{main} function of the benchmarking program with the command-line arguments defined by SPEC CPU2006.

\subsubsection{Microbenchmarks for quantification of system performance}

In this experiment, we implemented several microbenchmarks considered to be appropriate by qualitatively investigating programs of the target benchmarks.
Table \ref{tab:microbenchmarks} shows our implemented microbenchmarks.

\begin{table}[t]
	\centering
	\caption{Environments and compiler options.}
	\label{tab:environments}
	\begin{tabular}{ll}
		\Hline
		\textbf{OS} (kernel) & Linux 3.2.0-4-686-PAE \\
		\textbf{Android NDK} & Revision 10e (for Linux x86) \\
		\textbf{ \texttt{gcc} Version} (for x86-64) & 4.8  \\
		\textbf{ \texttt{gcc} Version} (for A32) & 4.9 \\
		\textbf{ \texttt{gcc} Version} (for A64) & 4.9 \\
		\textbf{ \texttt{gcc} Options} (excerpt) & \texttt{-O0 -std=c99} \\
		\Hline
	\end{tabular}
\end{table}

\begin{table*}[t]
	\centering
	\begin{threeparttable}[htb]
		\centering
		\caption{Microbenchmarks we implemented.}
		\label{tab:microbenchmarks}
		\begin{tabular}{p{2.2cm}p{13.5cm}}
		\Hline
		\textbf{Name} & \textbf{Program details} \\
		\hline
		${\rm loop}(n)$ & \small Process an empty loop composed of comparison instructions, branch instructions, and integer add instructions $2^n$ times.\\
		\hline
		${\rm INTadd}(k, n)$ & \small Process a loop containing $k$ integer add instructions without memory access $2^n$ times.\\
		\hline
		${\rm INTmul}(k, n)$ & \small Process a loop containing $k$ integer multiply instructions without memory access $2^n$ times.\\
		\hline
		${\rm INTaddmul}(k, n)$ & \small Process a loop containing interleaved execution of $k$ integer add instructions without memory access and $k$ integer multiply instructions without memory access $2^n$ times.\\
		\hline
		${\rm INTdiv}(k, n)$ & \small Process a loop containing $k$ integer division $2^n$ times\tnote{*1}.\\
		\hline
		${\rm INTstore}(k, n)$ & \small Process a loop containing $k$ integer store instructions which have the same destination virtual address.  $2^n$ times.\\
		\hline
		${\rm INTstoreload}(k, n)$ & \small Process a loop containing interleaved execution of $k$ integer store instructions and $k$ integer load instructions which have the same destination/source virtual address $2^n$ times.\\
		\hline
		${\rm FPadd}(k, n)$ & \small Process a loop containing $k$ floating point add instructions without memory access $2^n$ times\tnote{*2}.\\
		\hline
		${\rm FPmul}(k, n)$ & \small Process a loop containing $k$ floating point multiply instructions without memory access $2^n$ times\tnote{*2}.\\
		\hline
		${\rm FPdiv}(k, n)$ & \small Process a loop containing $k$ floating point divide instructions without memory access $2^n$ times\tnote{*2}.\\
		\hline
		${\rm FPstore}(k, n)$ & \small Process a loop containing $k$ floating point store instructions which have the same destination virtual address.  $2^n$ times\tnote{*2}.\\
		\hline
		${\rm FPstoreload}(k, n)$ & \small Process a loop containing interleaved execution of $k$ floating point store instructions and $k$ floating point load instructions which have the same destination/source virtual address $2^n$ times\tnote{*2}.\\
		\Hline
		\end{tabular}
		\begin{tablenotes}
			\footnotesize
			\item[*1] Integer division is not always executed by one instruction. For example, some of ARM processors do not have integer divide instructions and requires a function call to perform an integer division.
			Therefore we specially heightened the level of abstraction of the benchmarking program used in ${\rm INTdiv} $.
			\item[*2] A floating point here refers to a 64-bit double precision floating point defined in IEEE 754.
		\end{tablenotes}
	\end{threeparttable}
\end{table*}

Each of microbenchmarks is written in C and we compiled with the same environments and the same options shown in Table \ref{tab:environments} as the target benchmarks.
We used \texttt{register} modifiers so that the compiler conducts register allocation in the presence of \texttt{-O0} option.

\subsubsection{Quantification with compound benchmark score}

Scores generated by the microbenchmarks shown in Table \ref{tab:microbenchmarks} contains an influence of some overhead code such as comparison instructions and branch instructions.
Therefore we introduce a value calculated from several benchmark scores and we call it \textit{compound benchmark score}.
We used compound benchmark scores shown in Table \ref{tab:definitionofcompoundbenchmarkscores} for quantifying factors of system performance.

\begin{table*}[t]
	\centering
	\caption{Compound benchmark score for quantifying factors of system performance. }
	\label{tab:definitionofcompoundbenchmarkscores}
	\begin{tabular}{llp{9.6cm}}
		\Hline
		\textbf{Name} & \textbf{Calculating Formula} & \textbf{Interpretation} \\
		\hline
		C\_Ia & \footnotesize \parbox[c][0.8cm][c]{0cm}{} \mbox{ $ \displaystyle{ \frac{{\rm INTadd}(24, 27) - {\rm INTadd}(6, 27)}{24 - 6} \cdot \frac{2^{31}}{2^{27}} } $} & Execution time of $ 2^{31} $ integer add instructions. \\
		\hline
		C\_Im & \footnotesize \parbox[c][0.8cm][c]{0cm}{} \mbox{ $ \displaystyle{ \frac{{\rm INTmul}(16, 27) - {\rm INTmul}(4, 27)}{16 - 4} \cdot \frac{2^{31}}{2^{27}} } $} & Execution time of $ 2^{31} $ integer multiply instructions. \\
		\hline
		C\_Iam & \footnotesize \parbox[c][0.8cm][c]{0cm}{} \mbox{ $ \displaystyle{ \frac{{\rm INTaddmul}(24, 26) - {\rm INTaddmul}(6, 26)}{24 - 6} \cdot \frac{2^{31}}{2^{26}} } $} & Execution time of interleaved $ 2^{31} $ integer add instructions and $ 2^{31} $ integer multiply instructions. \\
		\hline
		C\_Id & \footnotesize \parbox[c][0.8cm][c]{0cm}{} \mbox{ $ \displaystyle{ \frac{{\rm INTdiv}(24, 26) - {\rm INTdiv}(6, 26)}{24 - 6} \cdot \frac{2^{31}}{2^{26}} } $} & Execution time of $ 2^{31} $ integer divisions. \\
		\hline
		C\_Is & \footnotesize \parbox[c][0.8cm][c]{0cm}{} \mbox{ $ \displaystyle{ \frac{{\rm INTstore}(24, 29) - {\rm INTstore}(6, 29)}{24 - 6} \cdot \frac{2^{31}}{2^{29}} } $} & Execution time of $ 2^{31} $ integer store instructions which have the same destination virtual address. \\
		\hline
		C\_Isl & \footnotesize \parbox[c][0.8cm][c]{0cm}{} \mbox{ $ \displaystyle{ \frac{{\rm INTstoreload}(16, 27) - {\rm INTstoreload}(4, 27)}{16 - 4} \cdot \frac{2^{31}}{2^{27}} } $} & Execution time of interleaved $ 2^{31} $ integer store instructions and $ 2^{31} $ integer load instructions which have the same destination/source virtual address. \\
		\hline
		C\_Fa & \footnotesize \parbox[c][0.8cm][c]{0cm}{} \mbox{ $ \displaystyle{ \frac{{\rm FPadd}(16, 27) - {\rm FPadd}(4, 27)}{16 - 4} \cdot \frac{2^{31}}{2^{27}} } $} & Execution time of $ 2^{31} $ floating point add instructions. \\
		\hline
		C\_Fm & \footnotesize \parbox[c][0.8cm][c]{0cm}{} \mbox{ $ \displaystyle{ \frac{{\rm FPmul}(16, 27) - {\rm FPmul}(4, 27)}{16 - 4} \cdot \frac{2^{31}}{2^{27}} } $} & Execution time of $ 2^{31} $ floating point multiply instructions. \\
		\hline
		C\_Fd & \footnotesize \parbox[c][0.8cm][c]{0cm}{} \mbox{ $ \displaystyle{ \frac{{\rm FPdiv}(16, 24) - {\rm FPdiv}(4, 24)}{16 - 4} \cdot \frac{2^{31}}{2^{24}} } $} & Execution time of $ 2^{31} $ floating point divide instructions. \\
		\hline
		C\_Fs & \footnotesize \parbox[c][0.8cm][c]{0cm}{} \mbox{ $ \displaystyle{ \frac{{\rm FPstore}(16, 29) - {\rm FPstore}(4, 29)}{16 - 4} \cdot \frac{2^{31}}{2^{29}} } $} & Execution time of $ 2^{31} $ floating point store instructions which have the same destination virtual address. \\
		\hline
		C\_Fsl & \footnotesize \parbox[c][0.8cm][c]{0cm}{} \mbox{ $ \displaystyle{ \frac{{\rm FPstoreload}(16, 27) - {\rm FPstoreload}(4, 27)}{16 - 4} \cdot \frac{2^{31}}{2^{27}} } $} & Execution time of interleaved $ 2^{31} $ floating point store instructions and $ 2^{31} $ floating point load instructions which have the same destination/source virtual address. \\
		\Hline
	\end{tabular}
\end{table*}

\subsubsection{Other conditions}

\noindent \textbf{Time measurement.}
We measured execution time by having each program call the standard POSIX function \texttt{clock\_gettime} around the target code which execution time we measure.
The clock we used is \texttt{CLOCK\_MONOTONIC}.

\vspace{1.5mm}

\noindent \textbf{Number of trials.}
Considering that a benchmark score can vary depending on the state of the system at the start of benchmarking,
we conducted the trial several times.
Each of target benchmarks is run 3 times and each of the microbenchmarks 5 times. % 英語では前節に出てくる動詞を省略して良かったはず

However, because breakdown of a benchmark score requires only one score, we must summarize the multiple benchmark scores into one benchmark score.
In this experiment, we calculated summary values by using the minimum value of the benchmark scores because any benchmark score in this experiment is an execution time.
This is because an execution time cannot incidentally diminish, that is, a measurement of execution time always generates positive errors and therefore a minimal value of benchmark scores is the sample which has a lowest error.

\vspace{1.5mm}

\noindent \textbf{Regression analysis.}
In this experiment, we conducted multiple linear regression analysis,
where the input variables are the score of ${\rm loop}(31)$ and the compound score shown in Table \ref{tab:definitionofcompoundbenchmarkscores}
and the output variable is a score generated by one of the target benchmarks.
Because a degree that a program utilizes each factor cannot be negative,
we used non-negative least squares (NNLS) methods for deciding the partial regression coefficients corresponding to factor utilization degrees.
The NNLS method is done with \texttt{nnls} package (Version 1.4) of R language.

\subsection{Experimental Results and Discussion}

\begin{figure*}[t]
	\centering
	\includegraphics[width=.80\textwidth]{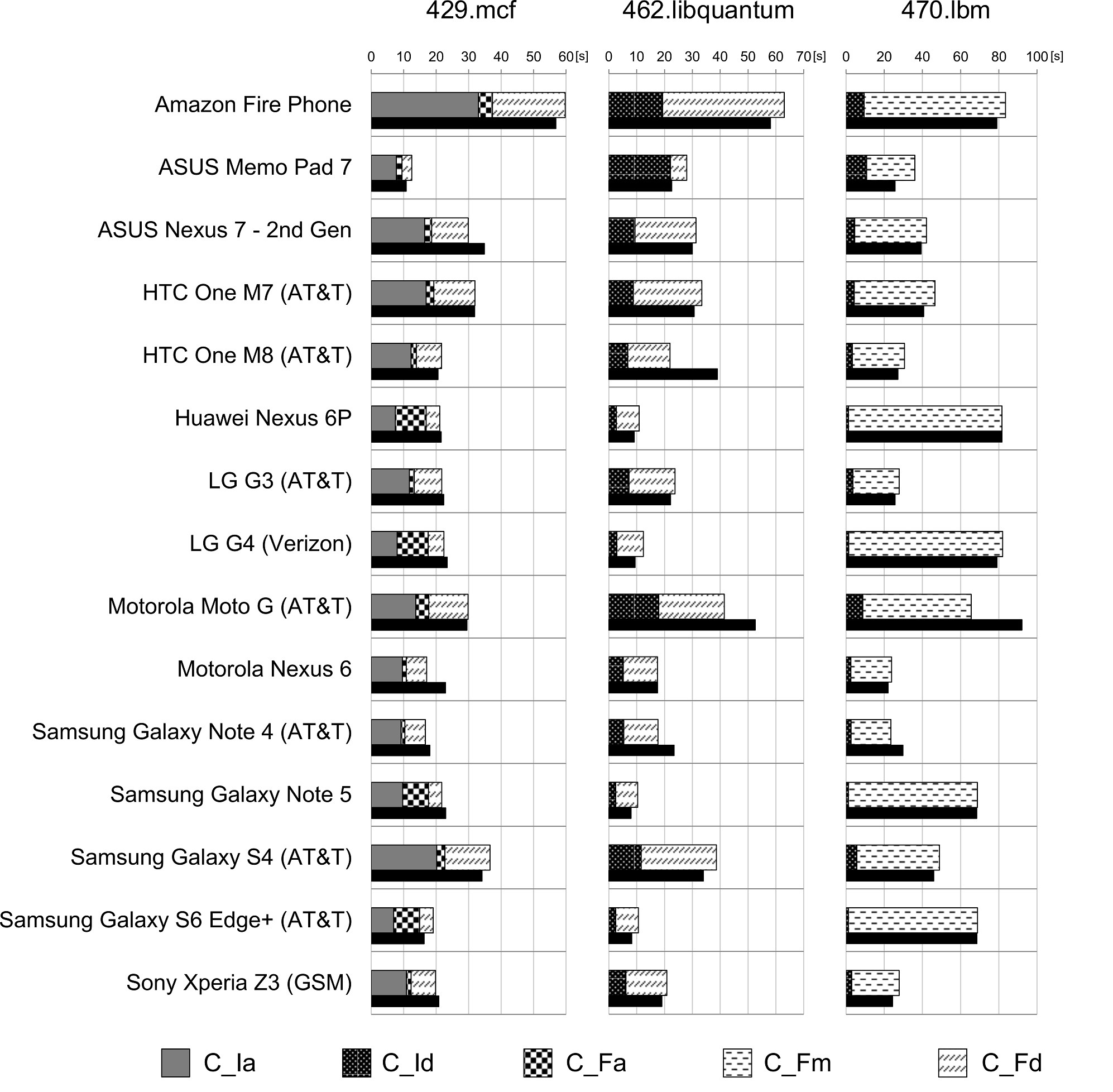}
	\caption{Result of score breakdown.}
	\label{fig:resultofcomponentanalysis1}
\end{figure*}

\noindent \textbf{Results.}
We show the result of breakdown of the 3 target benchmarks in Figure \ref{fig:resultofcomponentanalysis1}.
The black bars in the figure represent true values and the cumulative bars on the black bars represent scores estimated by the multiple regression analysis.

\vspace{1.5mm}

\noindent \textbf{Discussion.}
First, we discuss accuracy of the regression analysis.
While there is little difference between the true values and the estimated values in 429.mcf, 
there unignorable difference in 462.libquantum and 470.lbm. % 反復回避のための動詞省略では，任意の動詞を省略可能（参考: http://blog.livedoor.jp/y-46/archives/1257395.html）
This may be because the number of the microbenchmarks we used is not enough for multiple regression analysis.

Second, we discuss which factor contributes to the scores.

At a glance, we can notice that the only 5 of 12 factors contribute to the scores.
This may be because similar factors have been summarized into one factor due to their strong correlation.

We can also notice that the combinations of factor contributions in the scores of the 3 target benchmarks differ from each other.
This means our proposed method can distinguish differences between characteristics of benchmarking programs.

Meanwhile, in some of the target benchmarks, there is an abnormality in the combination of factor contributions.
Although C\_Ia (integer addition), C\_Fa (floating point addition), and C\_Fd (floating point division) contribute to the scores in 429.mcf and C\_Id (integer division) and C\_Fd (floating point division) to the scores in 462.libquantum,
this is considered to be an abnormal result because it is officially said that 429.mcf and 462.libquantum are benchmarks measuring integer arithmetic performance.
This may be because a contribution of integer arithmetic performance factor was mistaken for a contribution of floating point arithmetic performance factor due to their correlation.

\subsection{Summary}

This experiment showed that results of breakdown by our proposed method are sufficiently practical
and our proposed method can sufficiently distinguish differences between characteristics of benchmarking programs.
Meanwhile, we found that a result of breakdown can be worse in the form of an abnormal combination of factor contributions if microbenchmarks for quantifying system performance are not sufficiently appropriate.

\section{Conclusion}
\label{sec:conclusion}

In this paper,
we presented a method for breaking down a benchmark score without internal analysis of the benchmarking program.
The method quantifies a portion of number of aspects of the system performance with regression analysis using number of systems.
Experimental results with 3 benchmarks on 15 Android systems showed that our method could break down those benchmark scores even though there is room for improvement in accuracy.

Finally, we describe future work.
Although our proposed method heuristically prepares appropriate microbenchmarks and quantifies factors of system performance by using them,
our proposed method can be said to be still a na\"{i}ve method which has room for improvement because heuristics might have caused the accuracy degradation.
Possible improvements include the establishment of a method which (1) automatically collects candidates of appropriate benchmarks and (2) automatically selects an appropriate combination of those candidates.

Moreover, because it remains to be seen whether or not our proposed method can function well in another situation different from that of this experiment,
we need to perform some additional experiments using another system type, another benchmarks, and another regression analysis method.

\section*{Acknowledgments}

This work was supported by JSPS KAKENHI Grant Numbers 25700008 and 16K12406.
%This work was also supported by the New Energy and Industrial Technology Development Organization (NEDO) especially in a study on future directions such as utilization of a shared memory pool.

\bibliographystyle{IEEEtran}
\bibliography{IEEEabrv,myrefs}
%\begin{thebibliography}{99}% 文献数が10未満の時 {9}
%	\bibitem{}
%\end{thebibliography}

\end{document}